\documentclass[pss]{wiley2sp} % provides pss two-column style
\usepackage{bm}             
\usepackage{url}
\usepackage[utf8]{inputenc}
\usepackage[english]{babel}
\usepackage{graphicx}
\usepackage{subfig}

\usepackage{epstopdf}
\AppendGraphicsExtensions{.tif}

\begin{document}

% Title of the article
\title{Chemical-Strain Induced Tilted Dirac Nodes in (BEDT-TTF)$_2$X$_3$ (X = I, Cl, Br, F) Based Charge-Transfer Salts}

% Authors
\author{%
  R. Matthias Geilhufe\textsuperscript{\Ast,\textsf{\bfseries 1}},
  Benjamin Commeau\textsuperscript{\textsf{\bfseries 2}},
  Gayanath W. Fernando\textsuperscript{\textsf{\bfseries 2}},
%  Alexander V. Balatsky\textsuperscript{\textsf{\bfseries 1,2,3}}
}

%E-mail-address of corresponding author
\mail{e-mail
  \textsf{geilhufe@kth.se, gayanath.fernando@uconn.edu}}

% author's affiliations/addresses
\institute{%
  \textsuperscript{1}\,Nordita, KTH Royal Institute of Technology and Stockholm University, Roslagstullsbacken 23, SE-106 91 Stockholm, Sweden\\
  \textsuperscript{2}\,University of Connecticut, 2152 Hillside Road, U-3046 Storrs, CT 06269-3046, USA\\
%  \textsuperscript{3}\,Institute for Materials Science, Los Alamos National Laboratory, Los Alamos, NM 87545, USA}
}

% Please select about four verbal keywords for your manuscript.

\abstract{\bf%
The identification of novel multifunctional Dirac materials has been an ongoing effort. In this connection quasi 2-dimensional (BEDT-TTF)-based charge transfer salts are widely discussed. Here, we report about the electronic structure of $\alpha$-(BEDT-TTF)$_2$I$_3$ and $\kappa$-(BEDT-TTF)$_2$I$_3$ under a hypothetical substitution of iodine with the halogens bromine, chlorine and fluorine. The decreasing size of the anion layer corresponds to applying chemical strain which increases tremendously in the case of (BEDT-TTF)$_2$F$_3$. We performed structural optimization and electronic structure calculations in the framework of density functional theory, incorporating, first, the recently developed strongly constrained and appropriately normed semilocal density functional SCAN, and, second, van der Waals corrections to the PBE exchange correlation functional by means of the dDsC dispersion correction method. In the case of $\alpha$-(BEDT-TTF)$_2$F$_3$ the formation of over-tilted Dirac-type-II nodes within the quasi 2-dimensional Brillouin zone can be found. For $\kappa$-(BEDT-TTF)$_2$F$_3$, the recently reported topological transition within the electronic band structure cannot be revealed.}

\maketitle  

\section{\label{sec:level1}Introduction}

Organic charge transfer salts, especially quasi two-dimensional (BEDT-TTF)$_2$X$_3$ where X represent a variety of acceptor molecules, have generated  great interest over the past few decades. Depending on the structural phase, the acceptor molecules, and other factors, these can be associated with a rich variety of physical phenomena such as charge ordering, metal-insulator transitions, unconventional superconductivity, antiferromagnetism and spin liquids. In these compounds, the molecular units consisting of carbon atoms (i.e., donors) usually preserve their specific features and are separated by layers of anion atoms that act as acceptors. It is believed that the unpaired electrons (or holes) in the carbon $\pi$-orbitals are responsible for the conducting properties. 
%Some of these compounds undergo superconducting transitions under ambient pressure with the $\kappa$-phase producing the highest transition temperature of T$_C$=12 K. Other related compounds such as $\kappa$-(BEDT-TTF)$_2$Cu$_2$CN$_3$ have been discussed with regard to magnetic frustration and quantum spin liquids \cite{yshimizu2003}. (BEDT-TTF)$_2$X$_3$ salts show a certain degree of flexibility in packing motifs even for the same anion. For example, (BEDT-TTF)$_2$I$_3$ has been synthesized in five different structural phases, the $\alpha$-, $\beta$-, $\kappa$-, $\theta$- and $\lambda$-phases \cite{seo2004toward,endres1986x}; out of these, the $\alpha$-, $\beta$-, and $\kappa$-phases have attracted the majority of the attention. The $\alpha$-phase consists of stacks arranged in a herring-bone structure while the $\beta$-phase stacks are similar to the so-called Bechgaard salts showing parallel (face-to-face) 2-dimensional arrangements. The $\kappa$-phase consists of interacting dimer-type stacking.

The $\alpha$-phase with X=I represents a narrow band-gap semiconductor at ambient pressure \cite{tajima2006electronic} showing charge ordering as investigated by synchrotron X-ray diffraction measurements \cite{kakiuchi2007charge}. Under high pressure, it was reported that $\alpha$-(BEDT-TTF)$_2$I$_3$ exhibits a transition to a semi-metallic phase exhibiting tilted Dirac-crossings within the band structure \cite{liu2016insulating,hirata2016observation,kajita2014molecular,morinari2014possible,miyahara2014possible,kondo2009crystal,mori2009requirements}. Although Dirac materials have been of major interest recently \cite{wehling2014dirac}, $\alpha$-(BEDT-TTF)$_2$I$_3$ under high pressure is one of the few non-planar organic Dirac materials known to date \cite{geilhufe2017three,geilhufe2016data}. In general, organic compounds satisfying specific filling-constraints \cite{watanabe2016filling} favorable for exhibiting a semimetallic phase tend to be susceptible to interaction instabilities which break the global symmetry of the system and with that eliminate the protection of the semi-metallic phase \cite{wieder2018}. However, for sufficiently high temperatures, above the transition temperature of the instability, and incorporating chemical strain we argue that the quasi 2-dimensional material $\alpha$-(BEDT-TTF)$_2$F$_3$ is likely to host heavily tilted Dirac-cones at ambient pressure. 

We have recently reported first principles-based band structure calculations for $\alpha$- and $\beta$-phases with X=I and $\kappa$-phase with X=I, Br, Cl and F \cite{commeau}. For the latter it was found that the isovalent replacements of Iodine with Bromine, Chlorine and Fluorine leads to a chemical-strain induced topological band crossing along the path $\Gamma$-Y within the Brillouin zone, which is protected by the nonsymmorphic symmetry present for the $\kappa$-phase.

In the present work, we complement the previous study by  revising and extending this discussion for the $\kappa$- and $\alpha$-phases incorporating very recently developed exchange correlation functionals within the framework of the density functional theory (DFT). Here, $\alpha$-(BEDT-TTF)$_2$X$_3$ is found to be of particular interest, due to emerging tilted (and over-tilted) Dirac crossings within the band structure.  Furthermore, we rigorously discuss the structural changes and charge transfer connected to the incorporated chemical strain. We point out that the chemically induced strains may not necessarily be similar to those due to hydrostatic pressure.  Although the numerical values of charge transfer may depend on the selected appropriation scheme, the trends in charge transfer yield useful and consistent insights into the donor-acceptor nature of various atoms based on the DFT framework.

% The electronic structure calculations of the different phases was \cite{kobayashi1987crystal,mori2009requirements}. The influence of uniaxial strain and the charge ordering in the $\alpha$-phase were also widely investigated using ab-initio methods, e.g. by Kino and Miyazaki \cite{kino2006first} or Alemany \textit{et al.} \cite{alemany2012essential}. Correlation effects of the electrons, e.g., to describe the charge ordering in the $\alpha$-phase were discussed in terms of the Hubbard model \cite{Tanaka2016}. In our previous publication~\cite{commeau}, we presented electronic structure calculations for $\alpha$-, $\beta$- and $\kappa$-(BEDT-TTF)$_2$I$_3$ incorporating information about the symmetry of the states in terms of the irreducible representations. This information is important for the discussion of the gap closing in the high-pressure phase of  $\alpha$-(BEDT-TTF)$_2$I$_3$ and protection of line-nodes occurring on the Brillouin zone boundary of $\kappa$-(BEDT-TTF)$_2$I$_3$. 

The paper is structured as follows. First, we discuss the change of the cell volume by inducing chemical strain mediated by the substitution of the iodine layer by bromine, chlorine and fluorine. Second, we report about the changing charge transfer observed by the chemical substitution. Afterwards, we discuss the change of the electronic structure for $\alpha$-(BEDT-TTF)$_2$X$_3$ and $\kappa$-(BEDT-TTF)$_2$X$_3$ due to the decreased unit cell size. 

\section{Volume and Lattice Parameter Ratios Under Chemical Substitution}
We have examined the effects of changing the anion X in $\alpha$-(BEDT-TTF)$_2$X$_3$ and $\kappa$-(BEDT-TTF)$_2$X$_3$ in the framework of the DFT by applying a pseudopotential projector augmented-wave method~\cite{hamann1979norm,blochl1994projector,pseudo1,pseudo2}, as implemented in the Vienna Ab initio Simulation Package (VASP)~\cite{vasp2,kresse1999ultrasoft} and Quantum Espresso \cite{qespresso}. The exchange correlation functional was approximated by the recently developed semilocal meta-GGA functional (SCAN) \cite{sun2015,sun2016accurate}. Additionally, we performed calculations in the framework of  the generalized gradient approximation (PBE)~\cite{perdew1996}, incorporating Van der Waals corrections in terms of the dDsC dispersion correction method (VdW) \cite{steinmann2011comprehensive,steinmann2011generalized}. 
% \begin{figure}[b!]
% \includegraphics[width=8cm]{bedtttf.pdf}
% \caption{Crystal structure for different structural phases of (BEDT-TTF)$_2$I$_3$ viewed along the long axis of the BEDT-TTF molecules}
% \label{crystal_structure}
% \end{figure}

 \begin{figure}[h!]
\subfloat[]{\includegraphics[width=3.5cm]{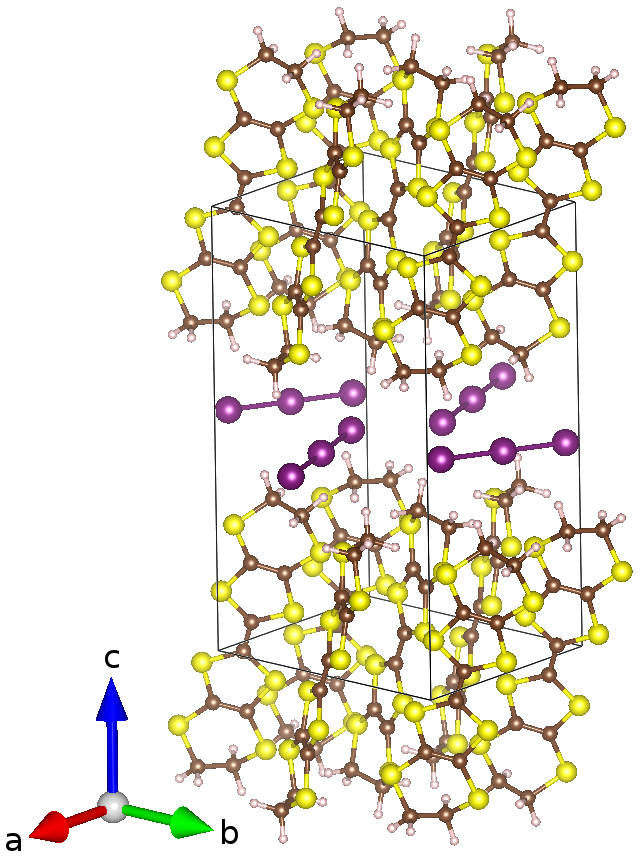}}\hfill
\subfloat[]{\includegraphics[width=4.cm]{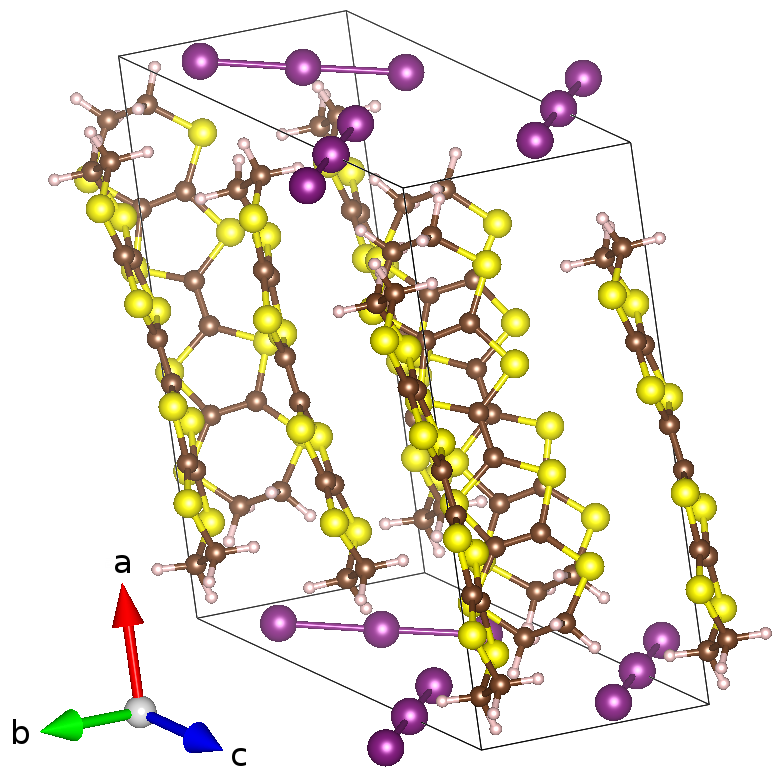}}\\
 \subfloat[\label{vol_change1}]{\includegraphics[width=7.1cm]{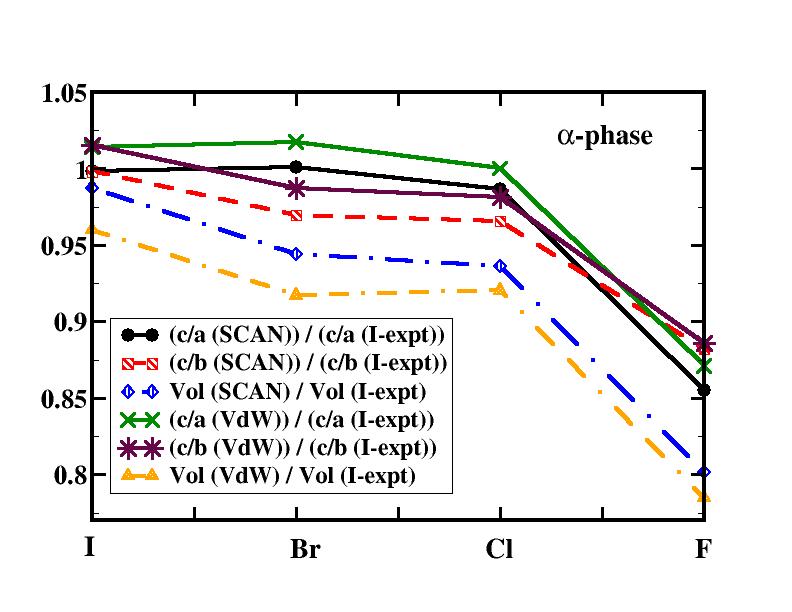}}\\
 \subfloat[\label{vol_change2}]{\includegraphics[width=7.1cm]{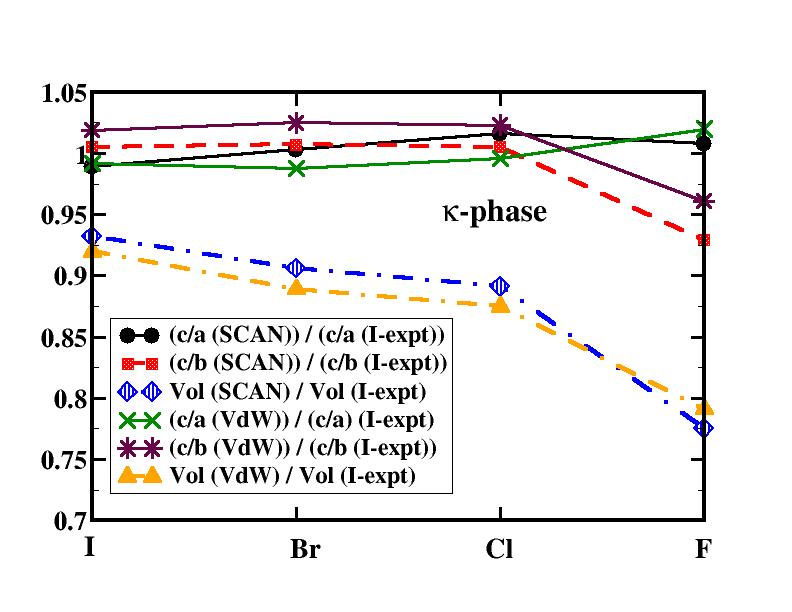}}
 \caption{The crystal structures with the respective $a$-, $b$-, and $c$-axes are shown in (a) for $\alpha$-(BEDT-TTF)$_2$X$_3$ and (b) for $\kappa$-(BEDT-TTF)$_2$X$_3$. Volume, $c/a$, and $c/b$ ratio changes under chemical substitution I$\rightarrow$ Br, Cl, Fr (c) $\alpha$-(BEDT-TTF)$_2$X$_3$ and (d) $\kappa$-(BEDT-TTF)$_2$X$_3$.}\label{ratios}
 \end{figure}
The calculations were performed without spin-orbit coupling, which is a reasonable approximation due to the light elements within the (BEDT-TTF)$_2$I$_3$ molecule. For the integration in $\vec{k}$-space, a $6\times6\times6$ $\Gamma$-centered mesh according to Monkhorst and Pack~\cite{monkhorst1976special} was chosen during the self-consistent cycle. The precision flag was set to ``normal''. A structural optimization was performed by allowing the ionic positions, the cell shape, and the cell volume to change ($\textsc{isif}=3$). 

Quantum ESPRESSO was applied to estimate the associated irreducible representations of the energy levels within the band structure. The cut-off energy for the wave function was chosen to be 48~Ry and the cut-off energy for the charge density and the potentials was chosen to be 316~Ry. The charge transfer for the $\kappa$-phase was computed using projected wavefunctions onto orthogonalized atomic wavefunctions, which calculates the Lowdin charges.

The basic structural parameters of $\alpha$-(BEDT-TTF)$_2$I$_3$ and $\kappa$-(BEDT-TTF)$_2$I$_3$ were taken from the Cambridge Structural Database (CSD) \cite{Groom:bm5086,endres1986x,kobayashi1987crystal}. An illustration of the molecular ordering as well as the chosen unit cells for $\alpha$-(BEDT-TTF)$_2$X$_3$ and $\kappa$-(BEDT-TTF)$_2$X$_3$ is shown in Figure \ref{ratios} (a), (b). The (BEDT-TTF) layers and the anion layer are stacked alternately in these materials and the latter is thought to be insulating. The bands near the Fermi level are mostly originating from (BEDT-TTF) layers, with the main contributions originating from the C and S $p$-orbitals \cite{commeau}. In the $\kappa$-phase, there is a set of bonding (occupied) and antibonding (partially occupied) bands in the vicinity of the Fermi level. We have examined the role of the anion layer as follows. The $\alpha$- and $\kappa$-phases of (BEDT-TTF)$_2$I$_3$ were studied by replacing all the iodine atoms  with another halogen atom, namely bromine, chlorine, and fluorine. These atoms are isovalent and smaller in size compared to iodine and hence can induce ``chemical strains", giving rise to a compression of the original unit cell.

We have monitored the relaxed volumes of the unit cell under such substitutions and found a steady decrease in the volume of the unit cell under the progressive substitutions I$\rightarrow$Br$\rightarrow$Cl$\rightarrow$F. Figures \ref{ratios} (c) and (d) show the relative change of the unit cell volume and $c/a$ and $c/b$ ratios with respect to the experimentally observed values \cite{endres1986x,kobayashi1987crystal} for $\alpha$-(BEDT-TTF)$_2$I$_3$ and $\kappa$-(BEDT-TTF)$_2$I$_3$.
In the $\alpha$-phase, under the above progressive substitution, $c/a$ and $c/b$ ratios also show an almost steady decrease.
In this case, the $c$-axis is the long axis of the crystal, while ${\hat a}$ and ${\hat b}$ vectors span a plane almost perpendicular to the $c$-axis. The steady decrease of the $c/a$ and $c/b$ ratios is due to a relatively large contraction along the $c$ axis compared to somewhat smaller changes in the plane spanned by ${\hat a}$ and ${\hat b}$ vectors. This change can be directly attributed to the decrease of the atom size in the insulating (acceptor) layers  under the substitution I$\rightarrow$Br$\rightarrow$Cl$\rightarrow$F.
The induced chemical strain may be designated as an almost uniaxial strain. 
The analogous changes in the $\kappa$-phase look more subtle. Note here that the long axis in this case is the $a$-axis while the $b$-axis is the special axis in this monoclinic structure (which is perpendicular to both $a$- and $c$-axes), according to
our choice of axes shown in Fig.~\ref{ratios} (d). Note that the $c/a$ ratio remains steady during the above progressive substitution
although both $a$ and $c$ parameters decrease. Also noteworthy is the fact that the parameter $b$ does not show changes comparable to
those associated with $c$ and $a$. This case may be interpreted as indicating a biaxial strain. 
The results are consistent for both exchange correlation functionals SCAN and VdW corrected PBE. For the $\alpha$-(BEDT-TTF)$_2$I$_3$, the structural information obtained incorporating the SCAN functional is in almost perfect agreement with the experimental values, whereas the cell volume is underestimated by about 5 \% using the VdW method. However, both functionals underestimate the cell volume for $\kappa$-(BEDT-TTF)$_2$I$_3$ by about 7\% for SCAN and 8\% for VdW. 

\section{Charge Transfer}
\begin{figure}[t!]
 \subfloat[Kappa Phase]{\includegraphics[width=8cm]{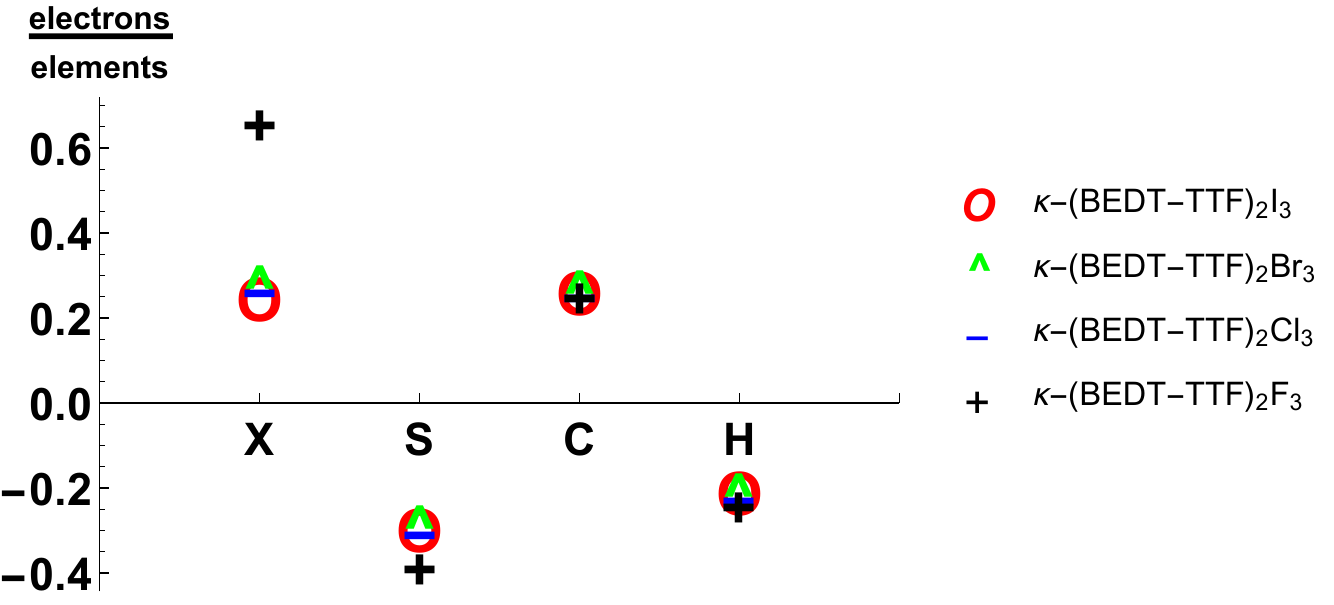}\label{kappa-charge-transfer-I-Br-Cl-F}}\\
 \subfloat[Alpha Phase]{\includegraphics[width=8cm]{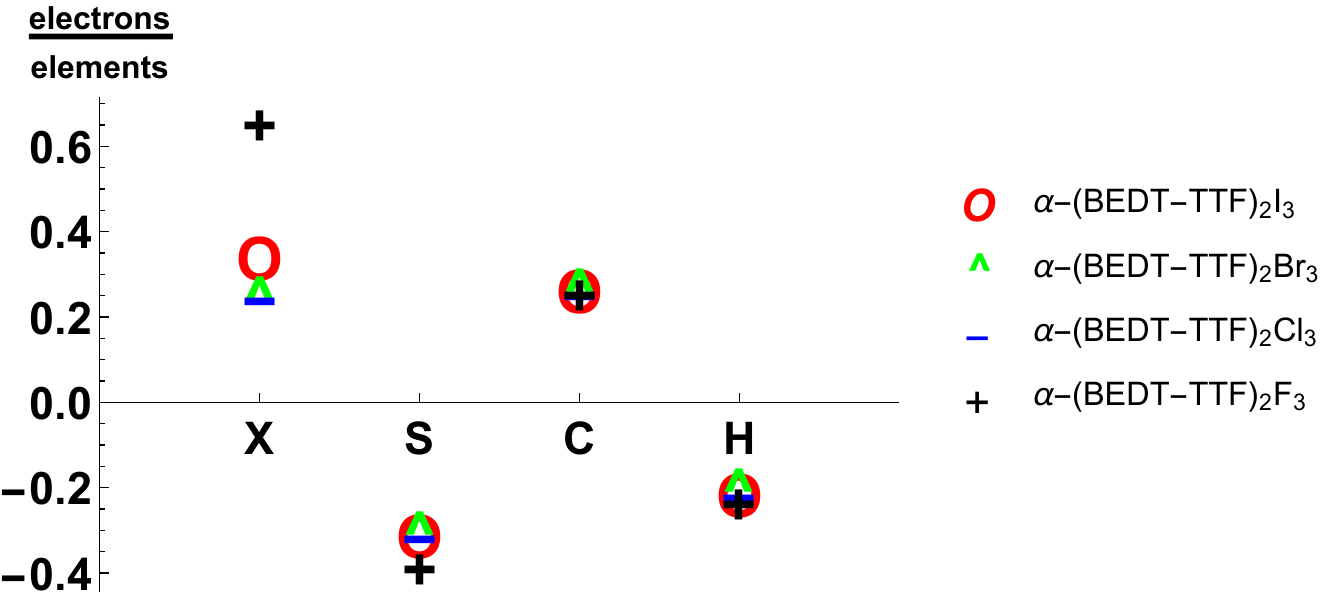}\label{alpha-charge-transfer-I-Br-Cl-F}}
 \caption{Charge transfer under chemical substitution for the halogen atoms (X=I,Br,Cl,F). The charge transfer is the net electron charge displaced with respect to the atom's valence electrons averaged over all atoms of the same element type. The valence electron numbers of each atom are (7,6,4,1) for (I,S,C,H) respectively. For example, $\kappa$-(BEDT-TTF)$_2$F$_3$ gained an additional 0.65 electron charge to each of its six fluorine atoms on average, giving it a total average of 7.65 electron charge to its valence electrons.}
 \end{figure}

Figure \ref{kappa-charge-transfer-I-Br-Cl-F} and \ref{alpha-charge-transfer-I-Br-Cl-F} illustrate the charge transfer for $\alpha$-(BEDT-TTF)$_2$X$_3$ and $\kappa$-(BEDT-TTF)$_2$X$_3$, respectively. Both sets of data have consistent charge transfer behavior. Charge is transfered from the sulfur and hydrogen atoms to the carbon and halogen atoms. Fluorine gains twice as much charge compared to the other halogens. 

\begin{figure}[h!]
  \subfloat[]{\includegraphics[width=3.6cm]{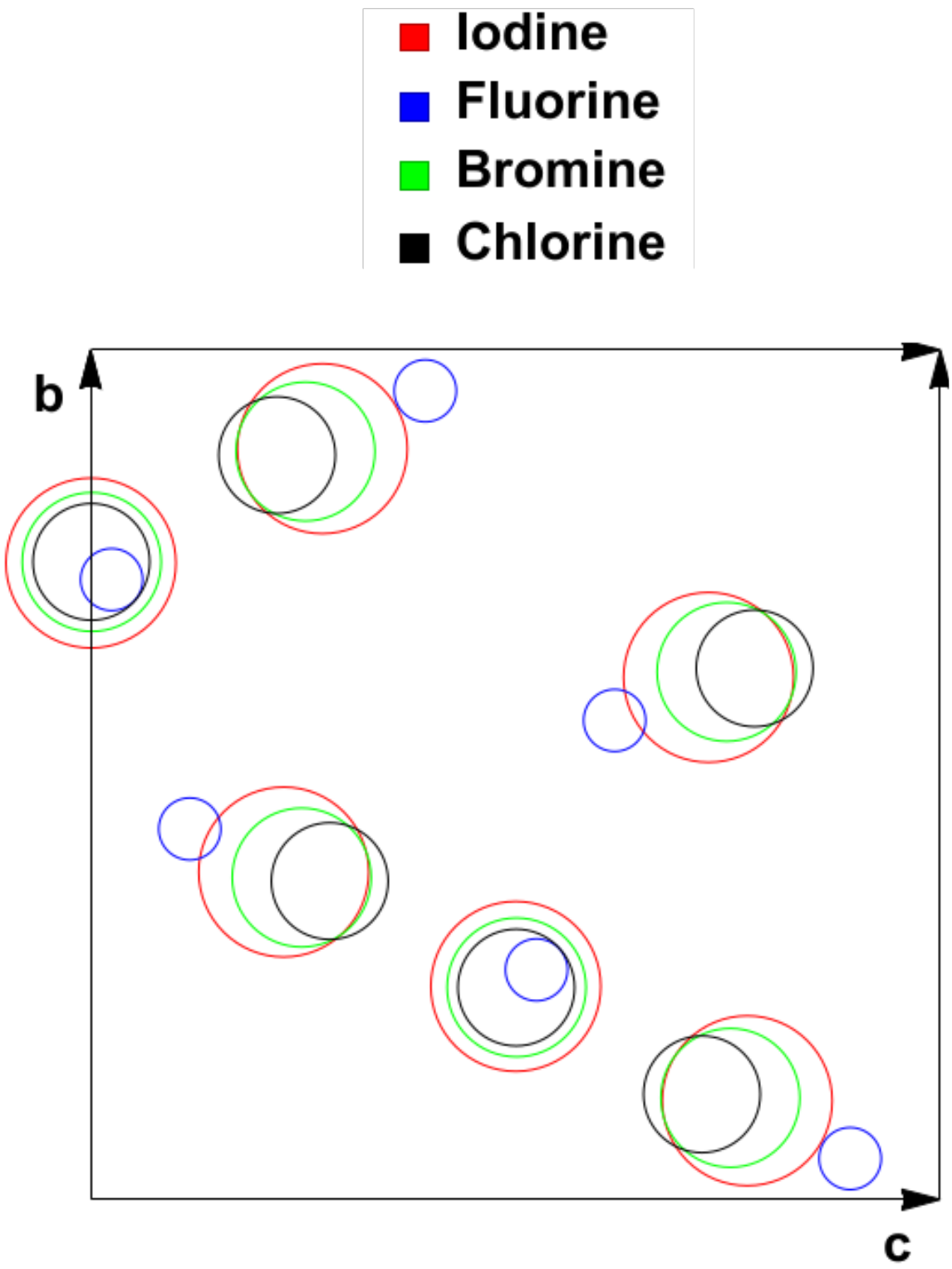}\label{halogen_movement}}\subfloat[]\hfill{\includegraphics[width=4.5cm]{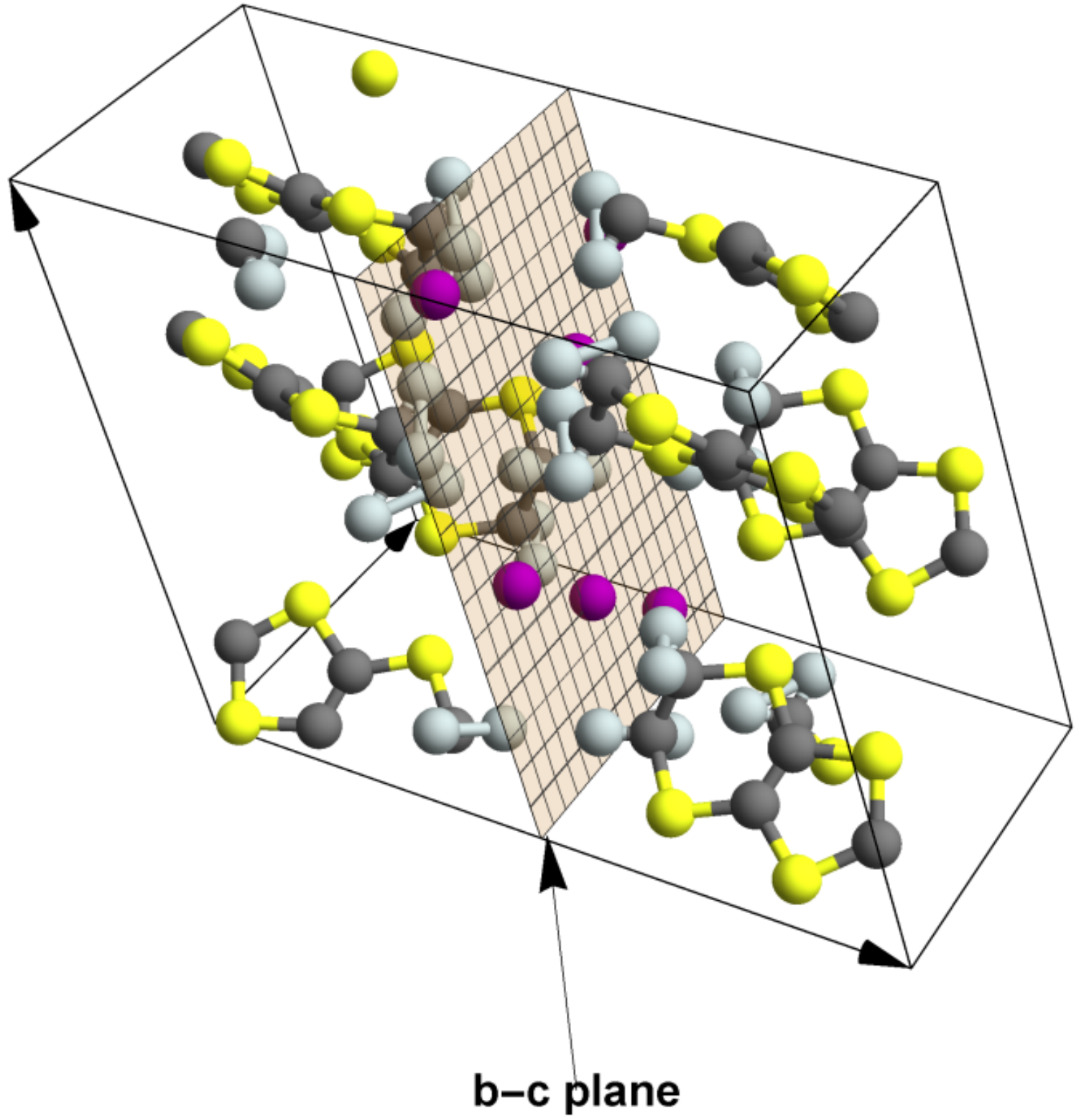}\label{BCplane}}
 \caption{Halogen movement under chemical substitution for $\kappa$-(BEDT-TTF)$_2$X$_3$ for X=(I,Br,Cl,F) in the $\vec{b}$-$\vec{c}$-plane, spanned by lattice vectors $\vec{b}$ and $\vec{c}$. (a) Top down view of normalized $\vec{b}$-$\vec{c}$-plane (lattice vectors normalized to equal length). Circles' centers are the halogen locations for  $\kappa$-(BEDT-TTF)$_2$X$_3$ for X=(I,Br,Cl,F). (b) $\vec{b}$-$\vec{c}$-plane cut where halogens (purple atoms) exist in 3D crystal unit cell for $\kappa$-(BEDT-TTF)$_2$X$_3$.}
 \end{figure}

Figure \ref{halogen_movement} illustrates a gradual movement of the halogens I$\rightarrow$Br$\rightarrow$Cl under chemical substitution ion relaxation in the kappa phase, and a large displacement in the location of 3 of its fluorine atoms. To our knowledge, there are no publications related to the synthesis of (BEDT-TTF)$_2$F$_3$. Fluorine, as a single atom, has shown in our relaxation calculations to be very electronegative and attempts to form strong charge transfer bonds with the other atoms in the crystal. However, we note here that organic crystals with an isolated anion fluorine atom can be synthesized in general\cite{touret2001crystal}.
 
\section{\label{sec:level2} Symmorphic $\alpha$-phase}
\begin{figure}[t!]
\subfloat[\label{bands_alpha_I}]{\includegraphics[width=8cm]{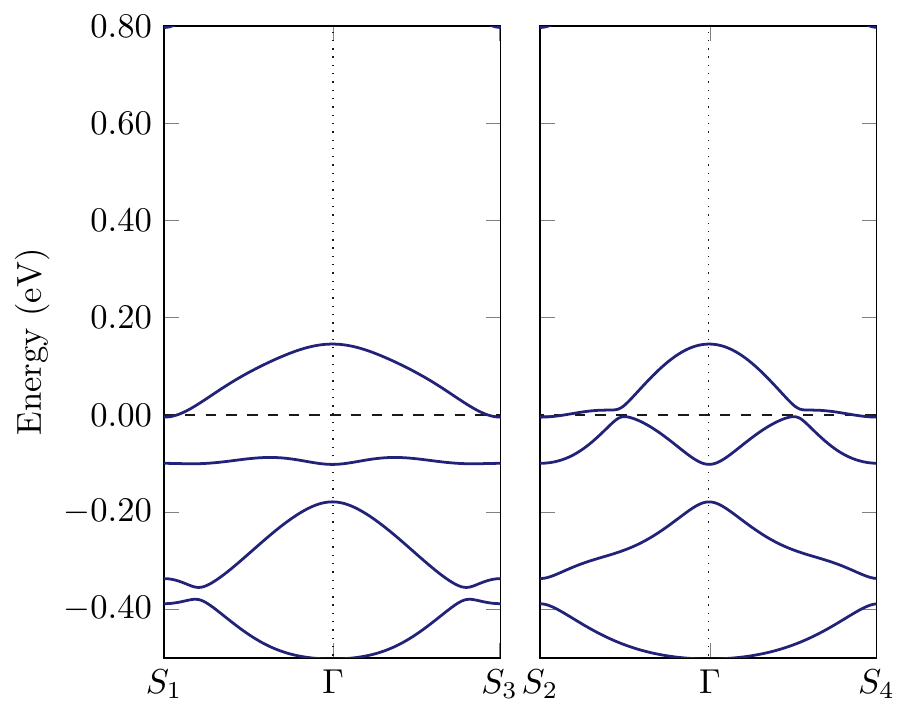}}\\
\subfloat[\label{bands_alpha_F}]{\includegraphics[width=8cm]{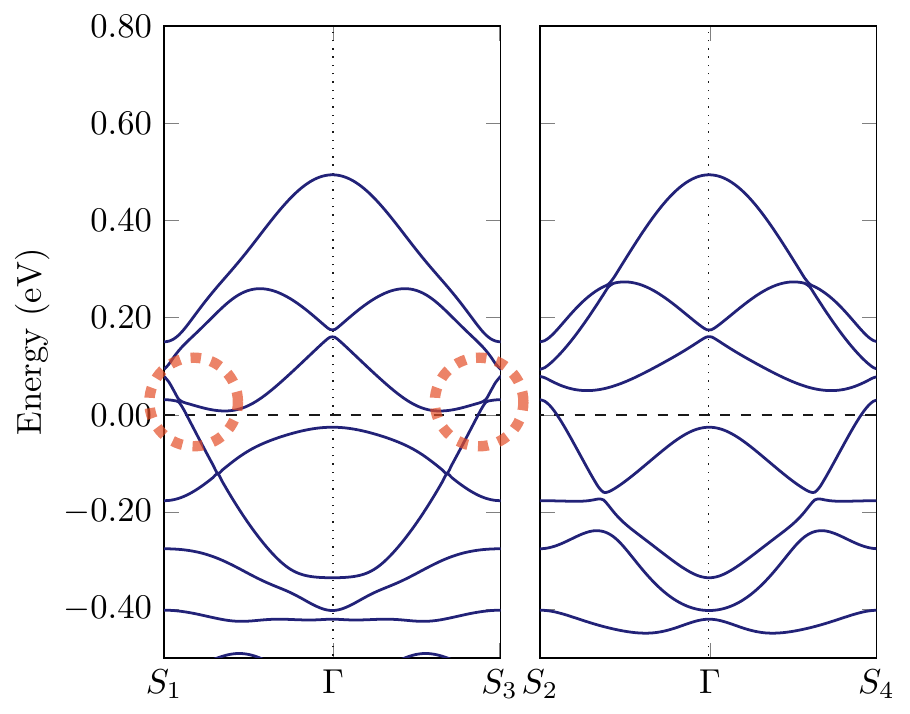}}
\caption{Band structure of (a) $\alpha$-(BEDT-TTF)$_2$I$_3$ and (b) $\alpha$-(BEDT-TTF)$_2$F$_3$ calculated using VASP and the SCAN meta-GGA functional. The high symmetry points represent the corners of the quasi-2D Brillouin zone given by $S_1=(0.5,0.5,0.0)$, $S_2=(0.5,-0.5,0.0)$, $S_3=(-0.5,0.5,0.0)$, and $S_4=(-0.5,0.5,0.0)$.}
\end{figure}

The $\alpha$-phases of (BEDT-TTF)$_2$X$_3$ crystallize in a triclinic crystal structure, having the (symmorphic) space group $P\overline{1}$ ($\#$2). 
%Since for every space group $(\mathcal{G}, \odot)$, the group of primitive lattice translations $\mathcal{T}$ forms a normal subgroup, the set of cosets $\mathcal{G}/\mathcal{T}$ forms a well defined group under coset multiplication, called the factor group, which is isomorphic to a point group. In the case of $P\overline{1}$, there are only two elements in the factor group and they are related to the identity $E$ and the inversion $I$,
%i.e., the corresponding space group, $\mathcal{G}$, expressed as a union of the cosets, is $\mathcal{G} = \mathcal{T}\cup (\left\{I|(0,0,0)\right\}\odot\mathcal{T})$. The irreducible representations (ireps), which are used to label the symmetries of the energy bands at any ${\vec k}$-point, are obtained from a factor group ${\mathcal {G}({\vec k}) }/{\mathcal{T}(\vec k)}$. Here $\mathcal {G}({\vec k})$ is a subgroup of $\mathcal G$ whose operations $\left\{R|{\vec t}\right\}$
%leave $R{\vec k} \cong\vec k$ (invariant to within a reciprocal lattice vector) while $\mathcal{T}(\vec k)$ consists of primitive translations $\left\{E|{\vec t_n}\right\}$ that satisfy $\exp(-i{\vec k}\cdot{\vec t_n})=1$. The structure of these groups determine the symmetries and irreps of the energy bands~\cite{bradley2010mathematical,HergertGeilhufe}.
Under ambient pressure, $\alpha$-(BEDT-TTF)$_2$I$_3$ is a narrow band-gap insulator as shown in Figure \ref{bands_alpha_I}. The ``tilted Dirac cone" reported in $\alpha$-(BEDT-TTF)$_2$I$_3$ under high pressure is centered at an interior k-point in the Brillouin zone \cite{liu2016insulating,hirata2016observation,kajita2014molecular,morinari2014possible,miyahara2014possible,kondo2009crystal,mori2009requirements}. The placement of the Dirac point in k-space is similar to what is observed in strained graphene where the Dirac points in unstrained graphene get dragged away from the high symmetry K and K$^{'}$ points. The ``tilt parameter" in this case has been shown to depend  linearly on the strain~\cite{anisotropic}. The linear (and tilted) bands, if placed near the Fermi level, can give rise to distinct transport properties. 

The substitution of I with Br and Cl, does not change the band structure qualitatively. However, the decreasing unit cell size lowers the energy of several high-lying bands and additionally decreases the band gap along the path $S_2-\Gamma-S_4$. Interestingly, the situation dramatically changes for a substitution $I\rightarrow F$, leading to a huge increase of the band gap along $S_2-\Gamma-S_3$ and over-tilted band crossings along $S_1-\Gamma-S_3$, as can be seen in Fig. \ref{bands_alpha_F}. As each band additionally carries a spin-degeneracy, the total degeneracy at the crossing points is four. Such materials are referred to as Dirac type-II semimetals. The same tilted crossings are revealed incorporating the PBE functional and Van der Waals corrections.

Recently, Dirac type-II and Weyl type-II materials have attracted a lot of attention, as they are discussed to effectively mimic the behavior of Dirac or Weyl fermions in the vicinity of a strong gravitational field \cite{Pyrialakos2017,Trescher2017,soluyanov2015type,volovik2016black}. The influence of the effective gravitational field leads to a curved metric and a tilted light or Dirac cone in the case of massless fermions. In this picture, an over-tilted Dirac-cone refers to Fermions beyond the event-horizon of a black hole, where the particle momentum is completely suppressed away from the origin of the gravitational field. Thus, the investigation of materials like $\alpha$-(BEDT-TTF)$_2$F$_3$ with effective excitations given by Dirac type-II fermions can lead to important insights of black hole physics by means of condensed matter systems. 

\section{Nonsymmorphic $\kappa$ phase}
The $\kappa$-phase crystallizes in a monoclinic crystal structure having the space group G $\equiv$ $P12_1 1$ ($\#$4), which can be represented by the coset decomposition
\begin{equation}
G = \mathcal{T} \cup (\left\{C_{2y}|(0,1/2,0)\right\}\odot \mathcal{T}).
\label{eq2}
\end{equation}
Here, $\left\{C_{2y}|(0,1/2,0)\right\}$ denotes a two-fold screw symmetry, represented by a two-fold rotation about the $y$-axis together with a non-primitive shift along the lattice vector $\vec{a}_2$. The nonsymmorphic nature of the space group leads to degenerate line-nodes along the Brillouin zone boundary. Nonsymmorphic symmetries were widely discussed in protecting Dirac nodes \cite{schoop2016dirac,wieder2016double,young2012dirac,chen2016topological}. In general, it is possible to distinguish between crossings protected by the crystalline symmetry, i.e., crossings which are associated with higher dimensional irreducible representations or pairs of complex-conjugate irreducible representations, and crossings protected by band topology, where the bands belong to different irreducible representations (accidental crossings), but where the crossing is forced to occur due to the connectivity of the bands \cite{geilhufe2017three,geilhufe2016data,PhysRevB.94.155108,bouhon2017global,Bradlyn2017}. 

\begin{figure}[t!]
\subfloat[\label{bands_kappa_I}]{\includegraphics[width=4.3cm]{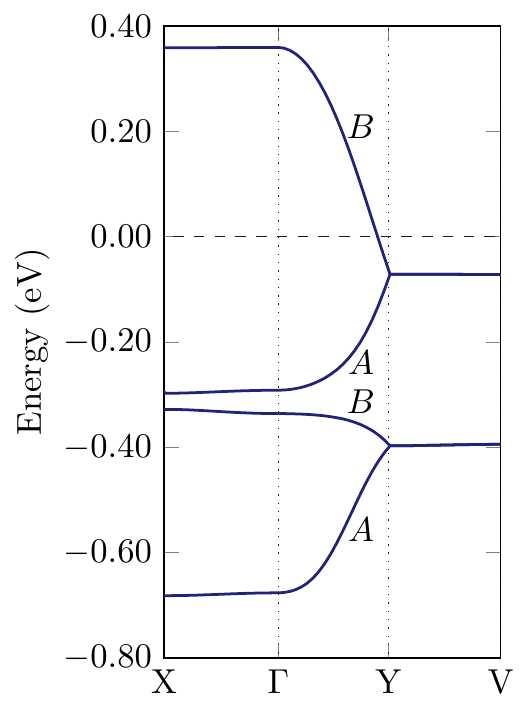}}
\subfloat[\label{bands_kappa_F}]{\includegraphics[width=4.3cm]{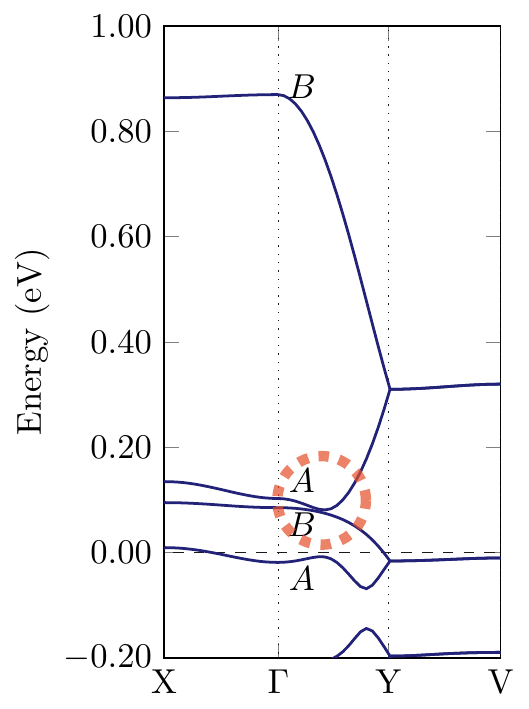}}
\caption{Band structure of (a) $\kappa$-(BEDT-TTF)$_2$I$_3$ and (b) $\kappa$-(BEDT-TTF)$_2$F$_3$ calculated using VASP and the SCAN meta-GGA functional. The high symmetry points are $X=(0.5,0.0,0.0)$, $Y=(0.0,0.5,0.0)$, and $V=(0.5,0.5,0.0)$.}
\end{figure}

The band structure of $\kappa$-(BEDT-TTF)$_2$I$_3$ close to the Fermi level is shown in Fig. \ref{bands_kappa_I}. Within the energy range of $[-0.8,0.4]$ two pairs of two bands can clearly be revealed. Here, the bands occur pair-wise as the nonsymmorphic symmetry protects the degeneracy at the Brillouin zone boundary. As discussed in \cite{commeau}, each pair consists of one band transforming even (irrep $A$ at the $\Gamma$-point) and one band transforming odd (irrep $B$ at the $\Gamma$-point) under $C_{2y}$. Here, the four bands illustrated in Fig. \ref{bands_kappa_I} correspond to an ordering of $A$, $B$, $A$, $B$, from the lowest to the highest band. However, in  Ref. \cite{commeau} it was reported that a substitution of $I\rightarrow F$ leads to a shift upwards of the bands, and, in addition to that, to a change of the ordering of the levels at the $\Gamma$ point to $A$, $A$, $B$, $B$. This ordering introduces a topologically protected crossing along the path $\Gamma-Y$. However, the present calculations performed using the SCAN functional do not support this change of the ordering, but reveal a  band touching along $\Gamma-Y$ (see Fig. \ref{bands_kappa_F}). A similar behavior is also found for the calculations performed using the VdW-corrected PBE functional.

\section{Universal Phase Diagrams}
 Universal phase diagrams summarizing the rich physics of the $\alpha$ and $\kappa$-phases have been presented following the
 work of Refs.~\cite{kanoda2,phase_diag}. The $\kappa$-phase diagram, which was published first in Ref.~\cite{kanoda2} for (BEDT-TTF)$_2$Y where
 Y=Cu(NCS)$_2$ combined with selected halogens, has been compared to the well-known (and strongly correlated) V$_2$O$_3$ phase diagram. It was noted there that, for this phase, the chemical pressure and the strength of correlations act in the opposite direction to the hydrostatic pressure. To our knowledge, the BEDT-TTF compound corresponding to fluorine substitution has not been synthesized. If it can be synthesized, its placement in such a phase diagram will be quite interesting. This is due to our calculated results which show a behavior somewhat different from the trends seen for Cl and Br substitutions. In the latter cases, the changes in the electronic and lattice structures are gradual and predictable. However, the changes in the lattice constant ratios as well as volume are relatively significant under fluorine substitution. The induced chemical strains are not necessarily isotropic and may not resemble those due to hydrostatic pressure. The phase diagram indicates that the fluorine compound is likely to lead to equally or more rich physical properties, if it can be synthesized. %We note here that since fluorine is far more reactive (electronegative) when compared to other halogens, the related compound may not be stable. 
 However, any other small set of atoms as the insulating layers (even though not chemically similar) may give rise to fascinating physical properties.
 %\begin{figure}[h!]
%\includegraphics[width=6cm]{kappa_phase.jpg}
%\caption{Universal phase diagram of the $\kappa$-phase - %from ref.\cite{phase_diag} - need to redo this %one.\label{kappa_phase}}
%\end{figure}

\section{Conclusions}
Organic materials based on (BEDT-TTF)$_2$X$_3$ (X=I, Br, Cl, F) show great potential for fine tuning specific band structures near the Fermi level due to their softness. Our research shows that applying chemical strain to the $\alpha$- and $\kappa$- phases creates significant changes in the volume and ratios of their lattice constants. From the structural optimization it follows that the mutual replacement of iodine with bromine, chlorine and fluorine corresponds to an almost uniaxial strain applied to the sample. The structural optimization also give insights into the accuracy of the recently developed exchange-correlation functionals SCAN and VdW-corrected PBE. For SCAN we find a very good agreement with the experimental unit cell for $\alpha$-(BEDT-TTF)$_2$I$_3$. However, for the $\kappa$-phase, both functionals underestimate the cell volume. As a result of the chemical strain, we observe over-tilted Dirac-type-nodes in the $\alpha$-phase for $\alpha$-(BEDT-TTF)$_2$F$_3$. These crossings exhibit a four-fold degeneracy. For the nonsymmorphic $\kappa$-phase, the existence of a topological transition has been examined and questioned throughout this work. This research highlights the promise and importance of tuning the band structure using chemical or other strains that could lead to creating multi-functional Dirac materials. Additionally, our calculations can be used as a starting point to probe strong correlations known to exist in such compounds.

\section{Acknowledgement}\label{acknowledgements}
We thank Dr. A. V. Balatsky for his support and helpful discussions concerning this study.
%This work is supported in part by the Institute for Materials Science at Los Alamos through the US Department of Energy, BES E3B7. 
We are grateful for support from the Swedish Research Council Grant No.~638-2013-9243, the Knut and Alice Wallenberg Foundation, and the European Research Council under the European Union’s Seventh Framework Program (FP/2207-2013)/ERC Grant Agreement No.~DM-321031. The authors acknowledge computational resources from the Swedish National Infrastructure for Computing (SNIC) at the High Performance Computing Center North (HPC2N), the High Performance Computing (HPC) cluster at the University of Connecticut, and the computing resources provided by the Center for Functional Nanomaterials, 
which is a U.S. DOE Office of Science Facility, at Brookhaven National Laboratory under Contract No. DE-SC0012704.

\bibliographystyle{pss}
\bibliography{mybib.bib}

% \newpage

% \section*{Graphical Table of Contents\\}
% GTOC image:
% \begin{figure}[h]%
% \includegraphics[width=5.5cm]{Alpha_F.tif}
% \caption*{%
% Band structure of $\alpha$-(BEDT-TTF)$_2$F$_3$ calculated using VASP and the SCAN meta-GGA functional. The high symmetry points represent the corners of the quasi-2D Brillouin zone given by $S_1=(0.5,0.5,0.0)$, $S_2=(0.5,-0.5,0.0)$, $S_3=(-0.5,0.5,0.0)$, and $S_4=(-0.5,0.5,0.0)$. Applying chemical strain from $\alpha$-(BEDT-TTF)$_2$I$_3$ into $\alpha$-(BEDT-TTF)$_2$F$_3$ at ambient pressure recreates the same titled Dirac nodes observed around 1 GPa of uniaxial pressure with no chemical substitution.
% }
% \label{GTOC}
%\end{figure}
\end{document}